\documentclass{IEEEtran}
\usepackage{amssymb}
\usepackage{amsmath}
\usepackage{epstopdf}
\usepackage{attachfile}
\usepackage[latin1]{inputenc}
\usepackage{caption}
\usepackage{subcaption}
\usepackage{mathabx}
\usepackage{setspace}
\usepackage{psfrag}
\usepackage{cite}
\usepackage{multirow}
\usepackage{wasysym}
\usepackage{graphicx,times,amsmath,balance,amsfonts,cite,pgf,tikz,pgffor}

\newtheorem{remark}{Remark}

\begin{document}
\title{Low-Dimensional Shaping for High-Dimensional Lattice Codes}
\author{\IEEEauthorblockN{Nuwan S. Ferdinand, \IEEEmembership{Student Member,~IEEE}, Brian M. Kurkoski, \IEEEmembership{Member,~IEEE}, Matthew Nokleby, \IEEEmembership{Member,~IEEE} and Behnaam Aazhang, \IEEEmembership{Fellow,~IEEE}}
\thanks{N. S. Ferdinand was with Centre for Wireless Communications, University of Oulu, Finland and he is currently with University of Toronto, Toronto, ON, Canada (e-mail: nferdinand@ece.utoronto.ca). B. M. Kurkoski is with Japan Advanced Institute of Science and Technology, Nomi, Japan (e-mail: kurkoski@jaist.ac.jp). M. Nokleby is with Wayne State University, Detroit, MI, USA (e-mail: matthew.nokleby@wayne.edu). B. Aazhang is with Rice University, TX, USA and he is also the Finnish Academy Distinguished Professor (FiDiPro) at the Center for Wireless Communication, University of Oulu, Finland (e-mail: aaz@rice.edu).}
\thanks{This work was in part supported by the Academy of Finland, Tekes, GETA, the JSPS Kakenhi Grant Number 26289119, and the US National Science Foundation: ECCS - 1547305; CNS - 1527811.}
\thanks{This work was presented in part at the IEEE Information Theory Workshop, Hobart, Australia, November 2014 and the IEEE Symposium of Information Theory, Barcelona, Spain, July 2016.}}

\maketitle
\begin{abstract}
We propose two low-complexity lattice code constructions that have competitive coding and shaping gains. The first construction, named \emph{systematic Voronoi shaping}, maps short blocks of integers to the dithered Voronoi integers, which are dithered integers that are uniformly distributed over the Voronoi region of a low-dimensional shaping lattice. Then, these dithered Voronoi integers are encoded using a high-dimensional lattice retaining the same shaping and coding gains of low and high-dimensional lattices. A drawback to this construction is that there is no isomorphism between the underlying message and the lattice code, preventing its use in applications such as compute-and-forward. Therefore we propose a second construction, called {\em mixed nested lattice codes}, in which a high-dimensional coding lattice is nested inside a concatenation of low-dimensional shaping lattices. This construction not only retains the same shaping/coding gains as first construction but also provides the desired algebraic structure. We numerically study these methods, for point-to-point channels as well as compute-and-forward  using low-density lattice codes (LDLCs) as coding lattices and $E_8$ and {Barnes-Wall} as shaping lattices. Numerical results indicate a shaping gain of up to $0.86$ dB, compared to the state-of-the-art of $0.4$ dB; furthermore, the proposed method has lower complexity than state-of-the-art approaches.

\end{abstract}

\section{Introduction}
\label{sec:intro}
Lattice codes can achieve the capacity of the AWGN channel \cite{erez:IT04,Ling:IT14}; they use the same real algebra as the AWGN channel; and they have algebraic structure that makes them suitable for physical layer network coding, compute-and-forward, and interference alignment, etc., \cite{nazer:IT11,Ferdinand:TWC15,Feng:IT13, ordentlich:IT12,hong:IT13}. These information-theoretic results rely on random constructions of high-dimensional pairs of ``good'' lattices: one lattice provides the coding gain for AWGN channel, and the other lattice provides the shaping gain.

Recent years have seen the development of practical, low-complexity lattice codes with good coding gain \cite{sommer:IT08,Sadeghi:IT06, dipietro:ISIT13,shalvi:IT11,yona:AL13}. However, in order for these high coding gain lattices to be put in practice, they should satisfy a power constraint. In the lattice domain, the power constraint is satisfied by selecting a set of coding lattice points that are in a specific shaping region. One approach for this task is to use discrete Gaussian shaping as pointed out in \cite{Ling:IT14,Palgy:IEEEI12}. Based on the concept of \cite{Ling:IT14}, the discrete Gaussian shaping has been used in \cite{Yan:IT15}. Another approach, named \emph{systematic shaping}, was proposed in \cite{naftali:ITW09}. Although systematic shaping alone does not provide shaping gains, it was suggested in \cite{naftali:ITW09}, to use systematic shaping together with trellis shaping \cite{forney:IT92T} or shell mapping \cite{Laroia:IT94} to obtain shaping gains.

Another way to perform shaping is \emph{nested lattice shaping}, in which the Voronoi region of a high-dimensional sublattice is used as the shaping region. One of the advantages of nested lattice shaping over other shaping methods is it retains the algebraic structure between the messages and the lattice codes \cite{nazer:IT11}. However, the use of a high-dimensional lattice for shaping is costly, as the complexity of shaping increases sharply with the dimension. Further, developing high-dimensional lattices with good shaping gain is a challenging task. For example, \cite{naftali:ITW09} proposes a nested lattice shaping scheme for low-density lattice codes using the suboptimal $\mathcal M$-algorithm, and this approach yields only $0.4$ dB of the possible $1.53$ dB shaping gain.

In this paper, we propose two low-complexity lattice codebook constructions that result in good shaping/coding gains. Our approach is to develop shaping schemes using low-dimensional shaping lattices and high-dimensional coding lattices. {The first} construction is named \emph{systematic Voronoi shaping}. In this construction, as the first step, we propose an efficient algorithm to simultaneously maps short {blocks} of integers to the Voronoi region of a low-dimensional lattice. These mapped points are {named} \emph{dithered Voronoi integers}. When we do not use the dither, these Voronoi points {result} in \emph{Voronoi integers}, that is, integers that are uniformly distributed over the Voronoi region. Low-dimensional lattices such as $E_8$, the {Barnes-Wall} lattice ($BW_{16}$), and {the Leech lattice} can be used for this step. The second step of this construction is to encode these dithered Voronoi integers using a high-dimensional coding lattice with high coding gain. This step is performed using \emph{systematic lattice encoding}, which is based on the concepts of systematic shaping \cite{naftali:ITW09}, but {generalized} to {parity check matrices with} non-unit diagonal elements and the use of a subtractive dither. {Systematic} lattice encoding is a technique for mapping information integers onto lattice points such that the lattice point is near the corresponding integer sequence. This technique can be applied to any coding lattice with a lower-triangular parity check matrix, such as LDLCs \cite{sommer:IT08}, LDA lattices \cite{dipietro:ISIT13}, etc. As systematic lattice encoding only needs shaped integers to obtain shaping gains, we detail an alternative technique to obtain shaped integers and this method is called \emph{non-uniform integers}, which is based on non-uniform signaling \cite{Kschischang:IT93}. Then, we numerically study the performance of our code construction using {LDLCs}, showing that it retains the shaping gains of the shaped integers and the coding gains of LDLCs. Numerical results {show} that systematic Voronoi shaping results in $0.86$ dB shaping gain with the use of $BW_{16}$ as the shaping lattice. As the codewords are shaped, the marginal distribution of codeword elements are no longer {uniform}, hence we develop an approximated maximum a posteriori iterative decoder for LDLCs that accounts for the marginal distribution of the shaped codewords.

An important application of lattice codes is compute-and-forward \cite{nazer:IT11}, in which multiple sources transmit messages to relays that estimate a linear combination of incoming messages. Practical implementations of compute-and-forward were proposed in \cite{Ferdinand:TWC15,Feng:IT13,hong:IT13}. Particularly, in \cite{hong:IT13}, low-complexity scaler quantizer is used for compute-and-forward, albeit it has $1.53$ dB shaping loss due to scalar quantizer.  A necessary condition for compute-and-forward  is an isomorphism between linear combinations of lattice codes and linear combinations of information integers. The first code construction does not exhibit this algebraic structure, hence it is not suitable for compute-and-forward. Therefore, we develop a second code construction, named {\em mixed nested lattice coding}. In this construction, shaping is provided by a series of low-dimensional lattices into which a high-dimensional lattice such as an LDLC is nested. This construction possesses the shaping and coding gains of the respective shaping and coding lattices with the same encoding/decoding complexity of the first construction. Further, we prove that this construction retains the necessary algebraic structure such that linear combination of lattice codes can be mapped to a modulo linear combination of integers. Hence, the mixed nested lattice code construction, described in this section,  not only has good shaping/coding gains for the point-to-point AWGN channel, but it also has an advantage over the first construction as it has the necessary algebraic structure for applications such as compute-and-forward . We show this construction has a self-dithering property, hence it is practically desirable. Self-dithering means that the codewords are approximately uniformly distributed over the shaping region, as if dithering has been added, without explicitly doing so.

\textbf{Notation}: Matrices and vectors are denoted by bold upper and lower case letters, respectively. The $i$th element of vector $\mathbf a$ is denoted by $a_{i}$ and the $(i,j)$th element of a matrix $\mathbf A$ is denoted by $a_{ij}$. The Gaussian distribution with  mean $m$ and variance $\sigma^2$ is denoted by $\mathcal{N} (m,\sigma^2)$. The $n$-dimensional integer and real fields are denoted by $\mathbb{Z}^n$ and $\mathbb{R}^n$, respectively. $\mathbb F_{p^l}^n$ denotes the $n$-dimensional field with size $p^l$. The probability density function (PDF) of $x$ is denoted by $p(x)$. The modulo summation is denoted by $\bigoplus$. $\mathrm{diag}(\mathbf A)$ denotes the diagonal matrix with $i$th diagonal element $a_{ii}$. The transpose operation is denoted by $(\cdot)^T$. $\lfloor{\cdot}\rceil$ denotes element-wise rounding to the nearest integer.

\section{System Model and Preliminaries}
\subsection{System Model}
\label{sec:system}
We consider an AWGN channel model. The source encodes input information $\mathbf b \in \mathbb Z^n$ to a lattice point $\mathbf x' \in \mathbb R^n$ and transmits over an AWGN channel. The received signal
\begin{align}
\label{eqn:received}
\mathbf y = r\mathbf x' + \mathbf z,
\end{align}
where $\mathbf z$ is Gaussian noise vector with per-dimension variance $\sigma_z^2$, and $r$ is the channel fading coefficient. In Section \ref{sec:mix}, we will also consider the AWGN multiple-access channel for purposes of compute-and-forward.

\subsection{Lattice Codes}
An $n$-dimensional lattice $\Lambda_n$ is a discrete additive subgroup of $\mathbb R^n$. Any lattice can be obtained by taking integer multiplication of basis vectors. Taking these basis vectors as columns, the \emph{generator matrix} $\mathbf G \in \mathbb R^{n\times n}$ is formed such that $\Lambda_n = \mathbf G \mathbb Z^n$. The inverse of generator matrix $\mathbf G$ is denoted by $\mathbf H$ and it is called the \emph{parity check matrix}. The \emph{shortest-distance lattice quantization} is denoted by $\mathcal Q_{\Lambda_n}(\mathbf x)$, which maps any point $\mathbf x \in \mathbb R^n$ to the nearest point $\lambda \in \Lambda_n$:
\begin{align}
\mathcal Q_{\Lambda_n}(\mathbf x) = \arg \max_{\lambda \in \Lambda_n} \|\mathbf x -\lambda\|.
\end{align}
Scaling a vector before quantization is equivalent to quantizing the non-scaled vector by a scaled version of the lattice:
\begin{align}
\label{eqn:scalar.quantizer}
\mathcal Q_{\Lambda_n}(\alpha\mathbf x) = \alpha\mathcal Q_{\frac{\Lambda_n}{\alpha}}(\mathbf x),
\end{align}
where $\alpha$ is any scalar. The modulo-lattice operation with respect to $\Lambda_n$ returns the quantization error:
\begin{align}
\mathbf x \bmod \Lambda_n = \mathbf x - \mathcal Q_{\Lambda_n}(\mathbf x).
\end{align}
The modulo operation satisfies the following scalar transformation
\begin{align}
\label{eqn:scalar.modulo}
[\alpha\mathbf x] \bmod \Lambda_n = \alpha\left[\mathbf x \bmod \frac{\Lambda_n}{\alpha}\right].
\end{align}
Let $\mathcal P_n$ denote the fundamental parallelotope (or fundamental parallelepiped) region \cite[p. 4]{Conway-1999} {of} $\Lambda_n$ {with respect to a basis $\mathbf G$}:
\begin{align}
\mathcal P_n=\{ \mathbf G \mathbf s  | 0 \leq s_i < 1\} \label{eqn:para},
\end{align}
where $s_i$ is the $i$th element of $\mathbf s$. There is a shifted parallelotope region for each point of $\Lambda_{n}$. Any point in $\mathcal P_n$ is in exactly one such region. The \emph{fundamental Voronoi region}, denoted by {$\mathcal V_n \subset \mathbb R^n$}, of {$\Lambda_n$} is the set of points that are closer to $\lambda=\mathbf 0$ lattice point than to any other lattice point. Let $\mathbf d$ be a random dither that is uniformly distributed over the fundamental parallelepiped region (or the fundamental Voronoi region). Then, $(\mathbf x-\mathbf d)\bmod \Lambda_n$ is uniformly distributed over $\mathcal V_n$ for {any} $\mathbf x \in \mathbb R^n$, \cite[Chap. 4.2]{zamir:book14}.

The volume of the fundamental Voronoi region is denoted by $V(\Lambda_n)=\mathrm{Vol}(\mathcal{V}_n)$ and it is equal to
\begin{align}
\label{eqn:vol}
V(\Lambda_n)=\mathrm{Vol}(\mathcal{V}_n) = |\text{det}(\mathbf G)|,
\end{align}
where $\text{det}(\cdot)$ denotes the determinant operation. {Let $\alpha$ be a scalar, then $V(\alpha \Lambda_n)=\alpha^n V(\Lambda_n)$}. The {\em second moment} of a lattice $\Lambda_n$ characterizes the average power of a random variable uniformly distributed across $\mathcal{V}_n$:
\begin{align}
\label{eqn:second.moment}
\sigma_{x}^2=\frac{1}{n} E\left[ \|\mathbf x\|^2\right]=\frac{1}{nV(\Lambda_n)}\int_{\mathcal{V}_n}\left \| {\mathbf{x}} \right \|^2 d\mathbf{x}.
\end{align}
The \emph{normalized second moment} (NSM) of $\Lambda_n$ is defined as:
\begin{align}
\label{eqn:nsm}
G(\Lambda_n) = \frac{\sigma_x^2}{V(\Lambda_n)^{\frac{n}{2}}}.
\end{align}
The shaping gain of $\Lambda_n$ is defined as:
\begin{align}
\label{eqn:shaping.gain}
\gamma=\frac{G(\mathbb Z^n)}{G(\Lambda_n)}.
\end{align}

\section{Systematic Voronoi shaping}
\label{sec:code.construction1}
In order to be used as a capacity approaching channel code for the AWGN channel, a lattice code needs two elements: a coding lattice with high coding gain, and a shaping method that satisfies a power constraint with high shaping gain. Hence, in Sec.~\ref{sec.voronoi} we propose a two-step lattice code construction, named \emph{systematic Voronoi shaping}, for AWGN channels that results in good coding and shaping gains. The first step is to uniquely map the information integers to the dithered Voronoi integers, which are points inside the Voronoi region of the shaping lattice. Then, we use systematic lattice encoding approach \cite{naftali:ITW09} to encode these dithered Voronoi integers using a high-dimensional lattice such that the codewords retain the shaping gain from the first step and the coding gain from the high-dimensional lattice. Then, we discuss an alternative method, for dithered Voronoi integers, using non-uniform signaling concepts \cite{Kschischang:IT93}. Next, we discuss the two steps decoding operation for systematic Voronoi shaping. Lastly, we numerically evaluate the shaping and coding gains of our proposed schemes.

\subsection{Systematic Voronoi shaping: Encoding \label{sec.voronoi} }
First, we describe the \emph{dithered Voronoi integers}, a method to encode integers to dithered integers that are inside the fundamental Voronoi region of a shaping lattice. This mapping is bijective. The key idea here is to shape relatively small blocks of information integers using a low-dimensional lattice, then to stack them to form a high-dimensional vector, which is then encoded to a high-dimensional lattice. First, we describe properties of the coding and shaping lattices, after which we provide the steps of the proposed mapping. Then, we encode the these concatenated points using a high-dimensional lattice. We show that resulted code construction approaches the same shaping and coding gains as the shaping and coding lattices.

{\bf \noindent Coding lattice: }
Let $\Lambda_{c,n}$ be the $n$-dimensional coding lattice defined by the lower-triangular parity check matrix $\mathbf H \in \mathbb R^{n\times n}$. Let $h_{ij}$ denote the $(i,j)$th element of $\mathbf H$. Let $\bar{\mathbf H}$ be the $n\times n$ diagonal matrix with $i$th diagonal element equal to $h_{ii}$, i.e., $\bar{\mathbf H}=\mathrm{diag}(\mathbf H)$. Let us divide the diagonal elements of $\mathbf H$ into $n/m$ groups where $m<<n$. Then we assume for the $r$th group, the diagonal elements are equal, i.e., $h_{ii}=h_r$ for $m(r-1)+1\leq i\leq rm$.

{\bf \noindent Shaping lattice: }
Let $\Lambda_{s,m}$ be the low $m$-dimensional lattice, let $\Theta \in \mathbb{R}^{m \times m}$ denote its generator matrix, let $\mathcal V_{s,m}$ be its fundamental Voronoi region, and $\mathcal P_m$ be its fundamental parallelepiped. The generator matrix $\Theta$ must satisfy several properties as follows. First, $\Theta$ is lower-triangular. Second, the diagonal entries of $\Theta$ scaled by any diagonal element of $\mathbf H$ must be integers, i.e.
\begin{align}\label{eqn:prime.power}
	h_{jj}\theta_{ii} \in \mathbb Z, \;\;\;\; \forall \ j=\{1,\ldots n\}.
\end{align}
Finally, for each column $j$,
\begin{equation}\label{eqn:integer.elements}
	\theta_{ij}/\theta_{jj} \in \mathbb{Z}, \;\;\;\; \forall \ i.
\end{equation}
Well-known lattices such as $D_m$, $E_8$, and $BW_{16}$, scaled by $h_{jj}^{-1}M$, $2h_{jj}^{-1}M$, and $\sqrt{2}h_{jj}^{-1}M$ satisfy these conditions for $M \in \mathbb Z$. These lattices have good shaping gains, and they also have low-complexity decoding algorithms \cite{Conway-1999}, which make them ideal for practical implementation.

{\bf \noindent Dithered Voronoi integers:}
Let the transmitter have an $n$-dimensional integer column vector $\mathbf b \in \mathbb Z^n$. First, it divides the integer vector $\mathbf b$ to $n/m$ integer vectors such that $\mathbf b=[(\mathbf b^1)^T, (\mathbf b^2)^T, \ldots (\mathbf b^{n/m})^T]^T$ where $\mathbf b^r \in \mathbb Z^m$.  Then, these integer vectors $\mathbf b^r$ are uniquely mapped to the points in fundamental Voronoi region of $h_r\Lambda_{s,m}$. This mapping is bijective.

The mapping procedure of integer information $\mathbf  b^r = (b_1^r, \ldots, b_m^r)^T$ to $\mathbf c^r = (c_1^r, \ldots, c_m^r)^T  \in \mathcal V_{h_r\Lambda_{s,m}}$ goes as follows. We approach this problem by first mapping the $\mathbf  b^r$ to integers inside the fundamental parallelepiped region of the scaled shaping lattice $h_r\Lambda_{s,m}$. Hence, we first choose $0 \leq b_i^r \leq h_r\theta_{ii}-1$, which yields the code rate of
\begin{align}
\mathcal R &= \frac 1 m \sum_{ii=1}^m\log_2 h_r\theta_{ii}  \textrm{ bits/dimension}.
\end{align}
We define $\mathbf f^r$ by normalizing each element by $h_r\theta_{ii}$:
\begin{align}
\label{eqn:scaled.information}
\mathbf f^r &= \left( \frac{b_1^r}{h_r\theta_{11}},  \frac{b_2^r}{h_r\theta_{22}}, \cdots ,  \frac{b_m^r}{h_r\theta_{mm}} \right)^T.
\end{align}
As an element of $\mathbf f^r \in (0,1]^m$, $h_r \Theta \mathbf f^r$ is in the fundamental parallelepiped region $h_r\mathcal P_m$ of the scaled shaping lattice $h_r\Lambda_{s,m}$ according to the definition of \eqref{eqn:para}. In general, $h_r\Theta \mathbf f^r$ is not a lattice point of $h_r\Lambda_{s,m}$. However, it is an integer vector by the properties of $\Theta$:
\begin{align}
h_r\Theta \mathbf f^r \in \mathbb Z^m.
\end{align}
It is also labeled by a unique $\mathbf b^r$. Now, we create an $m$-dimensional \emph{subtractive dither}\footnote{We say subtractive dither, when it is known to both transmitter and receiver \cite[Definition 4.1.1]{zamir:book14}.} vector that is uniformly distributed over $\mathcal P_m$. First we select the $i$th element $a_i^r$ that is uniformly distributed over $[0,\theta_{ii})$, then we find the subtractive dither vector that is uniformly distributed over $\mathcal P_m$:
\begin{align}
\label{eqn:dither1}
\mathbf d^r = \Theta \mathbf a^r.
\end{align}
Now we form an $n$-dimensional dither vector:
\begin{align}
\label{eqn:dither.n1}
\mathbf d =[(\mathbf d^1)^T \; (\mathbf d^2)^T\ldots (\mathbf d^{n/m})^T]^{T}.
\end{align}
As fundamental parallelepiped partitions the lattice space, we can find a representative point for $h_r\Theta \mathbf f^r- h_r \mathbf d_r$ inside the fundamental Voronoi region $\mathcal V_{h_r\Lambda_{s,m}}$:
\begin{align}
\label{eqn:encode}
\mathbf c^r - h_r \mathbf d_r &=  h_r\Theta \mathbf f^r- h_r \mathbf d_r - \mathcal Q_{h_r\Lambda_{s,m}}\left( h_r\Theta \mathbf f^r-h_r \mathbf d_r\right )
\nonumber \\ & = [h_r\Theta \mathbf f^r-h_r \mathbf d_r] \bmod h_r \Lambda_{s,m}.
\end{align}
where $\mathcal Q_{h_r\Lambda_{s,m}}\left( \cdot \right ) \in \mathbb Z^m$ based on \eqref{eqn:prime.power} and \eqref{eqn:integer.elements}, hence $\mathbf c^r \in \mathbb Z^m$. This mapping procedure uniquely maps the integer information $\mathbf b^r$, selected from $0 \leq b_i^r \leq h_r\theta_{ii}-1$, to a vector $\mathbf c^r - h_r\mathbf d_r $ that is inside the fundamental Voronoi region of the shaping lattice for a given dither $\mathbf d^r$. Then, by concatenating, we form the $n$-dimensional integer vector:
\begin{align}
\mathbf c =[(\mathbf c^1)^T \; (\mathbf c^2)^T\ldots (\mathbf c^{n/m})^T]^{T} \in \mathbb Z^n.
\end{align}

{\bf \noindent High dimensional lattice encoding:}

In this part, we present a framework based on \cite{naftali:ITW09} to encode Voronoi points to lattice points of a high-dimensional lattice while preserving the shaping gains. In \cite{naftali:ITW09}, this encoding framework is called {\em systematic shaping}.

In here, we generalize \emph{systematic lattice encoding}\footnote{We use the term ``systematic lattice encoding'' instead of ``systematic shaping'' to distinguish this method from the integer shaping procedure of the previous subsection.} for non-unit diagonal elements. By retaining the high coding gain properties of $\Lambda_{c,n}$, the systematic lattice encoding maps integer vectors to lattice points such that the integer vector elements can be obtained by simply rounding the lattice point coordinates. Systematic lattice encoding can be performed for any lattice if its parity check matrix is lower-triangular and the procedure is as follows.

Given $\mathbf c \in \mathbb Z^n$, we find $\mathbf x \in \Lambda_{c,n}$. This can be done by finding the integer vector $\mathbf k = (k_1,k_2, \ldots, k_n)^t$ such that
\begin{align}
\mathbf H \mathbf x &= (\mathbf c - \mathbf k) \textrm{\ and} \label{eqn.sysenc}\\ \nonumber
|h_{ii}x_i - c_i | &\leq \frac 1 2 \textrm{ for all $i=1, \ldots, n$}.
\end{align}
Note that line $i$ of \eqref{eqn.sysenc} is equivalent to
\begin{align}
h_{ii}x_i + \sum_{j=1}^{i-1}h_{ij} x_j = c_i - k_i.
\end{align}
Due to the triangular structure of $\mathbf H$, encoding is straightforward, with the $k_i$ and $x_i$ found recursively.
Clearly, $x_1 = c_1/h_{11}$ and $k_1 = 0$.  Continuing recursively for $i=2,3, \ldots, n$:
\begin{align}
\label{eqn:systematic.integer}
k_i=-\left \lfloor \sum_{j=1}^{i-1}h_{ij} x_j \right \rceil,
\end{align}
and
\begin{align}
x_i &= \frac{1}{h_{ii}} \left[c_i -\left(  \sum_{j=1}^{i-1}h_{ij} x_j - \left \lfloor \sum_{j=1}^{i-1}h_{ij} x_j \right \rceil \right)\right].
\end{align}
This encoding method guarantees that $|h_{ii}x_i -c_i| \leq 1/2$. After obtaining $\mathbf x$, we subtract the dither vector $\mathbf d$ to find the final lattice codeword:
\begin{align}
\mathbf x' &= \mathbf x - \mathbf d = \mathbf H^{-1}(\mathbf c - \mathbf k) - \mathbf d.
\end{align}
{\bf \noindent Shaping gain:}
Let us group $\mathbf x'$ into $n/m$ blocks as $\mathbf x' = [(\mathbf x'^1)^T \; (\mathbf x'^2)^T\ldots (\mathbf x'^{n/m})^T]^T$. Now we consider the $m$-dimensional $r$th block of the resulting lattice codeword $\mathbf x'$:
\begin{align}
 \mathbf x'^r  = h_r^{-1}(\mathbf c^r + \mathbf s^r) - \mathbf d^r = h_r^{-1}(\mathbf c^r - h_r\mathbf d^r + \mathbf s^r),
\end{align}
where $\mathbf s^r \in (-1/2,1/2)^m$. Now we substitute \eqref{eqn:encode}:
\begin{align}
\label{eqn:sys.code}
 \mathbf x'^r  &= h_r^{-1}\left([h_r\Theta \mathbf f^r - h_r \mathbf d^r] \bmod h_r \Lambda_{s,m}  + \mathbf s^r\right)
  \nonumber \\ & =[\Theta \mathbf f^r - \mathbf d^r] \bmod \Lambda_{s,m}  + h_r^{-1}\mathbf s^r.
\end{align}
According to \eqref{eqn:dither1}, $\mathbf d^r$ is uniformly distributed over fundamental parallelepiped $\mathcal P_m$, hence, based on the generalized dither concepts \cite[Chapter 4.2]{zamir:book14}, we can show that $[\Theta \mathbf f^r - \mathbf d^r] \bmod \Lambda_{s,m}$ is uniformly distributed over the Voronoi region of $\Lambda_{s,m}$, irrespective of the distribution of $\Theta \mathbf f^r$. Therefore, $[\Theta \mathbf f^r - \mathbf d^r] \bmod \Lambda_{s,m}$ has the same average second moment of $\Lambda_{s,m}$, hence, the average second moment of $\mathbf x'^r$ is
\begin{align}
\sigma_{\mathbf x'^r}^2 &= \frac{1}{n} E[\| {\mathbf x'}\|^2]  \nonumber \\ & = \frac{1}{n} E[\|[\Theta \mathbf f^r - \mathbf d^r] \bmod \Lambda_{s,m}  + h_r^{-1}\mathbf s^r\|^2]
\nonumber \\ & \leq\frac{1}{n} E[\|[\Theta \mathbf f^r - \mathbf d^r] \bmod \Lambda_{s,m}\|^2]  + \frac{1}{n} E[\|h_r^{-1}\mathbf s^r\|^2]
\nonumber \\ &  = \sigma_{\Lambda_{s,m}}^2 + h_r^{-2}\sigma_{\mathbf s^r}^2.
\end{align}
The volume of $\mathbf x'^r$ is $V_{x'^r} = V_{\Lambda_{s,m}}+V_s$ where $V_s$ corresponds to the addition volume due to $h_r^{-1}\mathbf s^r$ in \eqref{eqn:sys.code}. Based on this, NSM of $x'^r$ is
\begin{align}
\label{eqn:systematic.nsm}
G_{x'^r} &= \frac{\sigma_{x'^r}^2}{V_{x'^r}^{2/m}} \nonumber \\ & \geq \frac{\sigma_{\Lambda_{s,m}}^2 + h_r^{-2}\sigma_{\mathbf s^r}^2}{[V_{\Lambda_{s,m}}+ V_{s}]^{2/m}}  \nonumber \\ & = \frac{\sigma_{\Lambda_{s,m}}^2}{V_{\Lambda_{s,m}}^{2/m}}\left(1+\frac{V_{s}}{V_{\Lambda_{s,m}}}\right)^{-2/m} + \frac{h_r^{-2}\sigma_{\mathbf s^r}^2}{[V_{\Lambda_{s,m}}+ V_{s}]^{2/m}}
\nonumber \\ & = G_{\Lambda_{s,m}}\left(1+\frac{V_{s}}{V_{\Lambda_{s,m}}}\right)^{-2/m} + \frac{h_r^{-2}\sigma_{\mathbf s^r}^2}{[V_{\Lambda_{s,m}}+ V_{s}]^{2/m}}.
\end{align}
The higher the constellation size (or rate) that we use, the higher the $V_{\Lambda_{s,m}}$, e.g., if we use $ME_8$ as shaping lattice, then $V_{\Lambda_{s,m}}=M^mV_{E_8}$ and corresponding rate is $\mathcal R = \log_2(M)$ bits/dim. Hence, for large constellation sizes we can show that $G_{x'^r}$ approaches $G_{\Lambda_{s,m}}$ as $V_{s}$ and $h_r^{-2}\sigma_{\mathbf s^r}^2$ do not grow with constellation size. Therefore, it is evident from \eqref{eqn:shaping.gain} that the shaping gain of this encoding approaches that of $\Lambda_{s,m}$ for large constellation sizes. Numerical results verify this behavior of shaping gains in Sec. \ref{sec:numerical1}.

\remark The use of dither makes the elements uniformly distributed over the Voronoi region, hence, it gives the advantage of achieving the exact average second moment or the same shaping gain. When the quantization resolution increases, the role of dither becomes less prominent \cite[Chapter 4.1.1]{zamir:book14}. Let us consider an example. We use $ME_8$ as shaping lattice and $\mathbf c^r$ are integer vectors uniformly distributed over fundamental parallelepiped as obtained in \eqref{eqn:voronoi.shaped.mix}. The lattice quantizer is $\mathcal Q_{ME_8}(\mathbf c^r)=\mathcal Q_{E_8}(\mathbf x/M)M$, therefore for large constellations (large $M$), the number of possible points of $\mathbf c^r$ are large, hence, $\mathcal Q_{E_8}(\mathbf c^r/M)$ quantizer resolution is high. In other words, for large constellations, the distribution of $\mathbf c^r/M$ is approximately uniform over $\mathcal V_{E_8}$. Therefore, for large constellation sizes, the use of dither is less prominent in obtaining the same shaping gain. This fact is verified in numerical results. One can omit the use of dither depending on the practical application requirements.

\begin{figure}
    \centering
    \includegraphics[width=0.9\columnwidth]{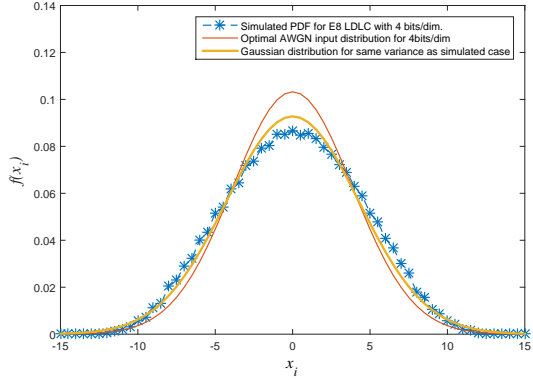}
    \caption{Marginal distribution of systematic LDLC Voronoi shaping.}
    \label{fig:systematic.voronoi}
\end{figure}

The marginal distribution of the codewords resulting from systematic Voronoi shaping output is plotted in Fig. \ref{fig:systematic.voronoi}. We have used the $E_8$ lattice to obtain Voronoi integers (we did not use the dither) and LDLC as the coding lattice. The rate is fixed to $4$ bits/dim and it is observed the output distribution is approximately Gaussian. The AWGN optimal input distribution is plotted for 4bits/dim, where it has an average power of $11.74$ dB. Use of the $E_8$ lattice with 4bits/dim results in average power of $12.67$dB and Gaussian distribution with average power of $12.67$ dB is plotted for comparison.

\remark
If the encoder has a set of integer vectors that have a certain shaping gain over the integer lattice, it is possible to get that shaping gain using systematic lattice encoding. It was suggested in \cite{naftali:ITW09} to use shell mapping \cite{Laroia:IT94} or trellis shaping \cite{forney:IT92T} to obtain shaped integers for this task. The shaping gains of Voronoi integers are comparable with the shell mapping for the same dimension, as can be seen in Table II \cite{Laroia:IT94}. However, the complexity of $m$-dimensional sphere shaped shell mapping is in the order of $\mathcal O (m^3\mathcal R^22^{2\mathcal R-1})$, and the storage requirement is on the order of $\mathcal O(m^2(\log m) \mathcal R 2^{2\mathcal R-1})$, where $\mathcal R$ is the code-rate. This shows the complexity is not only $3$rd order with the dimension but also depends on the rate. On the other hand, dithered Voronoi integers have linear complexity ($\mathcal O(m)$) with the dimension and complexity does not depend the rate. As discussed in \cite{Laroia:IT94}, the trellis shaping has several disadvantages compared to shell mapping including low shaping gains. Therefore, our proposed encoding is an alternative low-complexity method to shell mapping to be used with systematic encoding. Further, it has the advantage of implementing a dither, which is useful for low-rates.

{\bf \noindent Non-uniform integers--an alternative method:}
Here, we discuss an alternative method to obtain shaped integers. A key characteristic of the Voronoi mapping is that the resulting codewords are uniformly distributed over the shaping region. This is useful in practice because it results in fixed-rate transmission. However, the ultimate shaping gain for fixed-rate transmission is possible only with very high-dimensional lattices \cite{erez:IT04,erez:IT05}. If we relax the uniformity constraint, we can achieve near-optimal shaping gains even with small constellations \cite{Kschischang:IT93}. The following procedure stands as an alternative method to obtain shaped integers with the cost of variable-rate transmission.

Suppose a Bernoulli $1/2$ source. We map variable-length vectors of bits to integer vectors having a discrete Gaussian distribution. This is accomplished using the following procedure:

\begin{itemize}
  \item We first select the desired continuous Gaussian distribution. The variance of the distribution depends on the desired rate.
  \item We quantize the distribution to the integers and assign each integer its respective probability.
  \item Very low probability integers are omitted and the probability of each integer is normalized by sum probability.
  \item The Huffman procedure is performed using these integers and their probabilities to form a Huffman code dictionary. For each integer, this dictionary gives the unique bit vector. In $\mathtt{MATLAB}$, the command $\mathtt{huffmandict(integers,probabilities)}$ generates this dictionary.
  \item Finally, the variable length bits from source are assigned to respective integers based on Huffman code dictionary.
\end{itemize}
Then we form an $n$-dimensional vector of these integers, i.e. $\mathbf c \in \mathbb Z^n$ to use as an input to systematic lattice encoding as in the next step. As systematic encoding only slightly changes the average power, the shaping gain of the non-uniform integers is retained.

\subsection{Systematic Voronoi shaping: Decoding}
\label{sec:decoding}
This section proposes a two-step decoding scheme to recover the integer information from the received signal. The first step is to use the lattice $\Lambda_{c,n}$ to perform lattice decoding. The second step is to reverse the mapping from the Voronoi points to the integer information.

{\bf \noindent Lattice decoding using high-dimensional lattice:}
First, we add the scaled dither vector $r\mathbf H^{-1} \bar{\mathbf H} \mathbf d $ to the received signal \eqref{eqn:received}:
\begin{align}
\mathbf y' & = r\mathbf x' + \mathbf z +r \mathbf d
 \\\nonumber &= r\mathbf H^{-1} (\mathbf c-\mathbf k) + \mathbf z,
\end{align}
where $\mathbf H^{-1} (\mathbf c-\mathbf k)$ is a lattice point of $\Lambda_{c,n}$. In the next step, the decoder simply performs lattice decoding using the high-dimensional lattice:
\begin{align}
\label{eqn:lattice.decode}
\hat{\mathbf x} = \mathcal Q_{\Lambda_{c,n}}(\mathbf y'/r).
\end{align}
Then, the receiver performs element-wise rounding to find the respective integer vector:
\begin{align}
\label{eqn:estimte.integer}
\hat{\mathbf c}= \lfloor{\bar{\mathbf H} \hat{\mathbf x}}\rceil.
\end{align}
Although our code construction is general for any coding lattice with lower-triangular parity check matrix, in our numerical studies, we use LDLCs. Therefore, here we present a modified LDLC decoder that accounts for the input distribution. The standard LDLC decoder, proposed in \cite{sommer:IT08}, performs lattice decoding ignoring the shaping boundary. Hence, it ignores the marginal distribution of codeword elements, assuming they are equally likely. However, as we have noticed in Fig. \ref{fig:systematic.voronoi}, codeword elements are not equally likely. Here we propose an LDLC decoder that approximates MAP decoding according to the distribution over the codebook.
\begin{figure}
\centering
 \includegraphics[width=1\columnwidth]{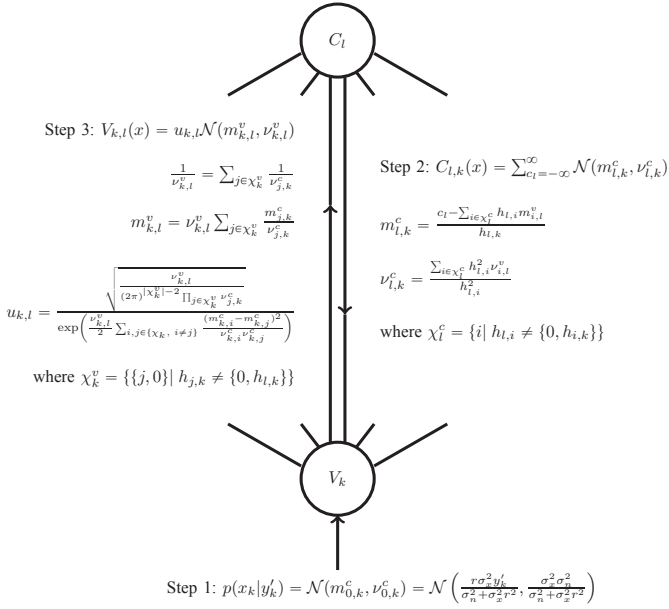}
 \caption{Approximated MAP LDLC message passing decoding algorithm.}
 \label{fig:bpldlc}
\end{figure}
Derivations of the following are based on several assumptions. First, we assume $x_i$ takes the Gaussian distribution given $\mathbf x$ is a lattice codeword.
Hence, we write:
\begin{align}
p(x_i|\mathbf x \in \Lambda_{c,n}) = \frac{1}{\sqrt{2\pi}\sigma_x} e^{-\frac{ x_i^2}{2\sigma_x^2}}, \; \; \forall i \in \{1,\ldots n\},
\end{align}
where $\sigma_x^2$ is the average power of lattice codeword $\mathbf x$. In order to calculate $p(x_i|\mathbf y')$, we use the same ``trick'' as in \cite[Sec. III]{sommer:IT08}, which assumes the elements $x_i$ are independent and identically distributed (i.i.d) with the necessary condition of $x \in \Lambda_{c,n}$. Due to the i.i.d assumption, we have $p(x_i|\mathbf y')=p(x_i|y_i')$. We first calculate the correlation coefficient ($\rho_{xy'}$) between $x_i$ and $y_i'$:
\begin{align}
\rho_{xy'} &= \frac{E[y_i'x_i] - E[y_i']E[x_i] }{\sqrt{E[{y_i'}^2]-E[y_i']^2}\sqrt{E[x_i^2]-E[x_i]^2} } \nonumber \\ & = \frac{ r \sigma_x}{\sqrt{ r^2 \sigma_x^2 +\sigma_n^2} }.
\end{align}
Then we find
\begin{align}
p(x_i |  y_i') &= \frac{1}{\sqrt{2\pi\frac{\sigma_x^2\sigma_n^2}{\sigma_n^2+\sigma_x^2r^2}}} \exp\left({\frac{-\left (x_i-\frac{r\sigma_x^2 y_i'}{\sigma_n^2+\sigma_x^2 r^2}\right)^2}{2\frac{\sigma_x^2\sigma_n^2}{\sigma_n^2+\sigma_x^2r^2}}}\right).
\end{align}
The modified LDLC decoder uses $p(x_i |  y_i')$ as the input, which takes into account the codebook distribution, instead of $p(y_i' |  x_i)$, which is used in original LDLC decoder \cite{sommer:IT08}. The decoder steps are shown in Fig.~\ref{fig:bpldlc}\footnote{It is noted that similar MAP decoding have been suggested for multiple input multiple output (MIMO) channels in \cite{yona:AL13}.}.

{\bf \noindent Voronoi-reverse mapping:}
Let us suppose the lattice decoder \eqref{eqn:estimte.integer} correctly estimates the integer vector $\mathbf c$. Then it divides $\mathbf c$ into $n/m$ blocks. The Voronoi-reverse mapping is the reverse mapping operation of $\mathbf c^r$ to the information vector $\mathbf b^r$, described as follows. By definition, $\mathcal Q_{h_r\Lambda_{s,m}}\left( \cdot \right )$ is a lattice point of $\Lambda_{s,m}$, hence we can represent it as $-\mathcal Q_{h_r\Lambda_{s,m}}\left( h_r\Theta \mathbf f^r-h_r \mathbf d_r\right ) = h_r\Theta \bar{\mathbf f}^r$, where $\bar{\mathbf f}^r \in \mathbb Z^m$. Then, using \eqref{eqn:encode}, any point $\mathbf c^r$ can be written as:
\begin{align}
\mathbf c^r &=h_r\Theta \mathbf f^r + h_r\Theta \bar{\mathbf f}^r , \label{eqn:decamp}
\end{align}
where the $i$th element of $\mathbf f^r$ is $0 \leq f_i^r < 1$ by definition \eqref{eqn:scaled.information}. Here $\mathbf c^r$ is in the parallelepiped for $\Theta \bar{\mathbf f}^r$. Using the lower-triangular structure of $\Theta$, the first row of \eqref{eqn:decamp} is:
\begin{align}
c_1^r &=  h_r\theta_{11} (\bar{f}_1^r + f_1^r),
\end{align}
which has a unique solution since $\bar{f}_1^r$ is an integer and $f_1^r$ is fractional. Continuing recursively for $i=2,3, \ldots, m$,
\begin{align}
\label{eqn:decode.voronoi.integer}
c_i^r &=   h_r\theta_{ii} (\bar{f}_i^r + f_i^r) + \sum_{j=1}^{i-1} h_r\theta_{ij} (\bar{f}_j^r + f_j^r),
\end{align}
it is always possible to find unique $\bar{f}_i^r$ and $f_i^r$. A decoding algorithm is given as follows:
\begin{enumerate}
\item Input: $\mathbf c^r$ with elements $c_i^r$ and generator matrix $h_r\Theta$ with elements $h_r\theta_{ij}$
\item For each $i=1,2, \ldots, m$:
\begin{enumerate}
 \item Let $t_i^r = \bar{f}_i^r + f_i^r $, then find it using \eqref{eqn:decode.voronoi.integer}:
 \begin{align}
t_i^r &= \frac{ c_i^r - \sum_{j=1}^{i-1} h_r\theta_{ij} t_j^r}{h_r\theta_{ii}},
\end{align}
\item find the integer part $\bar{f}_i^r$:
\begin{align}
\bar{f}_i^r &= \lfloor t_i^r \rfloor,
\end{align}
\item find the information integer $b_i^r$:
\begin{align}
b_i^r &= (t_i^r - \bar{f}_i^r) \theta_{ii}h_r .
\end{align}
\end{enumerate}
\item Output: integer vector $\mathbf b^r = (b_1^r, \ldots, b_m^r)$.
\end{enumerate}

\subsection{Numerical evaluation}
\label{sec:numerical1}
Efficient quantization (lattice decoding) schemes are available for $E_8$ and $BW_{16}$ lattices \cite{Conway:IT82}; further, $E_8, BW_{16}$ lattices have the best shaping gains among $8$ and $16$ dimensional lattices, which are found to be $0.65, 0.86$ dB \cite{Conway-1999}. Hence, we use $E_8$ and $BW_{16}$ lattices as shaping lattice to perform systematic Voronoi shaping. The LDLC is used as the high-dimensional coding lattice and LDLC parity check matrix given in \cite{naftali:ITW09} is used, which has unit-diagonal elements.

\begin{figure}
    \centering
    \includegraphics[width=1\columnwidth]{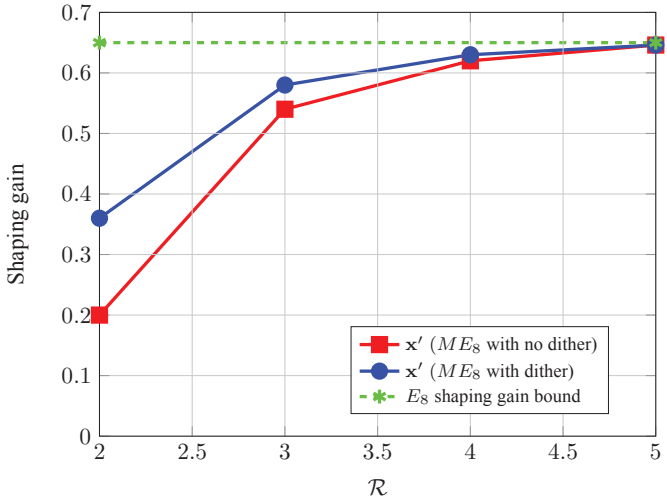}
    \caption{Shaping gain for scaled $E_8$ lattice with LDLC.}
    \label{fig:shaping}
\end{figure}

Fig.~\ref{fig:shaping} illustrates the shaping gains for systematic Voronoi shaping for different rates ($\mathcal R=\log_2(M)$) using $ME_8$ and LDLC lattices. When we do not use the dither, we observe shaping gains of $0.20,0.54,0.62$, and $0.65$ dB for constellations $M=4,8, 16$, and $32$. However, shaping gains of $0.36, 0.58, 0.63$ and $0.65$ dB are observed when we use the random dither. It is desired to use a fixed dither in practice. For the $E_8$ lattice, the best dither is \cite[Eq. 5]{Conway:IT83}:
\begin{align}
\label{eqn:optimal.dither}
\mathbf d_{\text{best}} = [ 0.01535 \;\;0.05002 &\;\;0.0831  \;\;0.14786 \;\;0.18069 \nonumber \\ &  \;\;0.21463  \;\;0.25040  \;\;0.71103 ],
\end{align}
and it results in the same shaping gains as using the random dither. Further, it is observed that the gap between the shaping gain of $\mathbf x'$ and shaping bound is significant for small constellation sizes irrespective of use of dither, however, it is less significant and asymptotically small for larger constellations. The reason for this gap is due to the fact that $ x'_i$ is uniformly distributed over $c_i \pm \frac 1 2$ and the effect of the additional $\frac 1 2 $ is significant for small constellations and for larger constellations it is less significant as discussed in \eqref{eqn:systematic.nsm}.

\begin{figure}
    \centering
    \includegraphics[width=1\columnwidth]{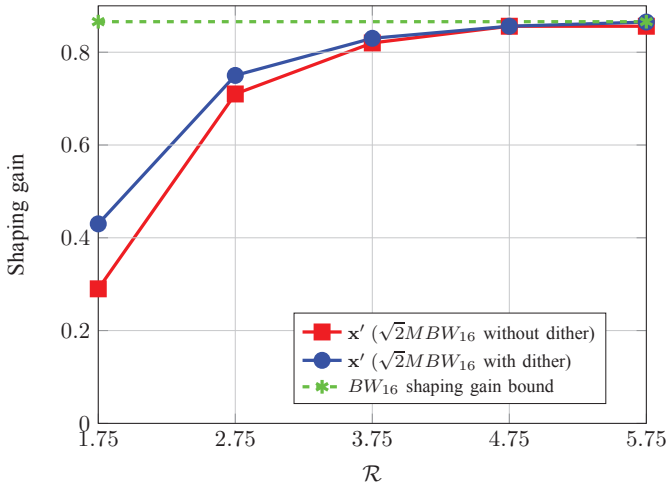}
    \caption{Shaping gain for scaled $BW_{16}$ lattice with LDLC.}
    \label{fig:BW.shaping}
\end{figure}%

We have simulated the $BW_{16}$ lattice to show the shaping gain performance in Fig.~\ref{fig:BW.shaping}. We observe similar behavior to $E_8$. As the constellation size increases, the shaping gain approaches that of the $BW_{16}$ shaping bound, which is approximately $0.86$ dB.

Therefore, from Fig.~\ref{fig:shaping} and Fig.~\ref{fig:BW.shaping}, we conclude that for small constellations (small rates), the use of dither (random or best) is important. Further, we conclude that the shaping gain approaches shaping bounds of shaping lattices as the constellation size becomes large, irrespective of dither been used.

\begin{figure}
    \centering
    \includegraphics[width=1\columnwidth]{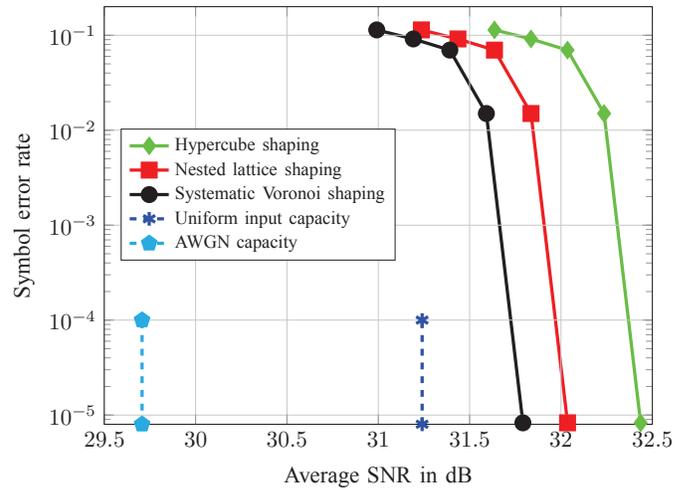}
    \caption{Symbol error rate versus average SNR for Voronoi integers. For $n=10^4$ and $\mathcal R=4.935$ bits/dimension.}
    \label{fig:ser}
\end{figure}

Fig.~\ref{fig:ser} shows the symbol error rate (SER) versus average SNR for systematic Voronoi shaping, with comparison to previously reported cases. The rate is fixed at $\mathcal R=4.935$ bits/dimension and the block length is $n=10^4$; the slight rate penalty is due to the selection of constellation sizes for different rows of LDLC parity check matrix to protect the unprotected integers as described in \cite{naftali:ITW09}. Observe that the Voronoi integer shaping, using $E_8$ as the shaping lattice, has a $0.645$ dB gain over hypercube shaping and $0.25$ dB gain over the high-complexity nested lattice shaping \cite{naftali:ITW09}. With $E_8$ Voronoi integer shaping, LDLCs are only $0.65$ dB away from the rates achieved by the uniform input distribution at SER=$10^{-5}$ for $n=10^4$, which is $1.53$ dB away from AWGN capacity. This shows that LDLC performs close to uniform input distribution even with the inherited LDLC coding loss of $0.8$ dB for $n=10^4$ and the rate penalty of $0.4$ dB due to unprotected integers.

\begin{figure}
    \centering
    \includegraphics[width=1\columnwidth]{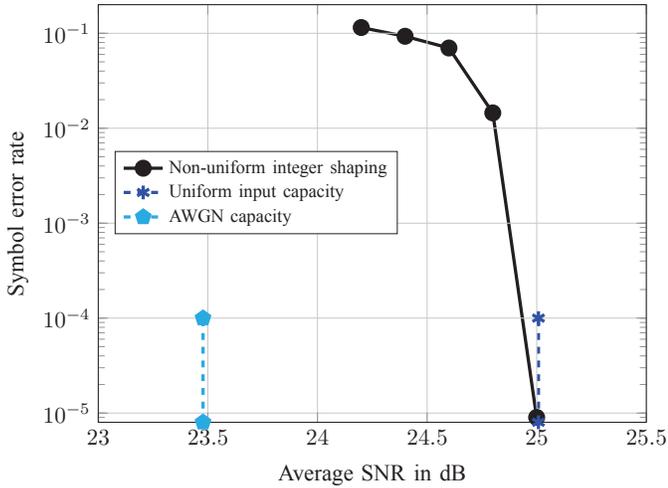}
    \caption{Symbol error rate versus average SNR for non-uniform integers. For $n=10^4$ and $\mathcal R=3.9028$ bits/dimension.}
    \label{fig:ser.non.uniform}
\end{figure}%

Fig.~\ref{fig:ser.non.uniform} shows the SER vs. average SNR for non-uniform integer shaping, with $n=10^4$. In order to protect the integer elements left less protected by the lower-triangular LDLC structure, three Gaussians are used for the Huffman procedure. For the first $9500$ elements, $\mathcal N(0,15)$ is used, for the second $350$ elements, $\mathcal N(0,15/5)$ is used, and $\mathcal N(0,15/9)$ is used for last $150$ elements. These distributions result in rates of $3.9675$, $2.7495$, and $2.4961$ bits/dimenation, respectively, and the average rate is calculated to be $(9500\times 3.9675+2.7495\times 350+2.4961\times 150)/10000=3.9028$ bits/dimension. Based on these integers, we obtain the LDLC code, then the second and last sets of codewords are protected by scaling with factors 2 and 4 respectively. Fig.~\ref{fig:ser.non.uniform} shows that at SER $10^{-5}$ non-uniform integer shaping shaping coincides with the uniform-input rate. Non-uniform integer shaping largely mitigates the shaping loss of LDLCs, leaving a $1.5$ dB gap to AWGN capacity due to the LDLC coding loss and additional loss due to less-protected integer elements.

\section{Mixed nested lattice codes}
\label{sec:mix}
In this section, we propose our second lattice code construction, named as ``mix nested lattice codes''. This construction preserves the algebraic structure, which is important in the recent work on lattice codes for multi-terminal applications. One such application is compute-and-forward  \cite{nazer:IT11}, in which multiple sources transmit messages to relays, and the relays estimate finite-field linear combinations of messages instead of the individual messages. Then the relays forward the estimated linear combination of messages. In this scenario, we must consider power constraints on both the transmitters' codewords and the relays' linear combination of codewords. We can enforce the latter constraint by making an explicit connection between integer combinations of lattice codewords and linear network coding over finite fields. A standard approach \cite{nazer:IT11} is to construct a Voronoi codebook $\mathcal C(\Lambda_{c,n}/\Lambda_{s,n})$ that is isomorphic to the finite field $\mathbb F_p^k$.

However, a codebook which is designed by systematic lattice encoding is not isomorphic to an underlying field. Hence, we cannot use the Sec.~\ref{sec:code.construction1} code constructions in scenarios such as two-way relays and physical-layer network coding, in which relays must forward linear functions of incoming messages.

Therefore, in this section, we propose mixed nested lattice codes as our second code construction and prove it has the necessary algebraic structure.  This construction creates a Voronoi codebook $\Lambda_{c,n}\cap \mathcal V_{s,n}$. It does not necessarily create the quotient nested lattice codebook $\Lambda_{c,n}/\Lambda_{s,n}$ as $\Lambda_{s,n}$ is not in general a sublattice of $\Lambda_{c,n}$. This construction can be used for {the} point-to-point AWGN channel as well for applications where algebraic structure is needed. Further, this construction has {a self-dithering property}, and hence is practically {appealing}.

\subsection{Mixed nested lattice codes}
\label{sec:new.com}
In this section we present {\em mixed nested lattice codes}, which uses distinct lattice pair to form a Voronoi codebook $\Lambda_{c,n}\cap \mathcal V_{s,n}$. The shaping lattice is constructed by {concatenating repetitions of} a low-dimensional lattice.  {A} high-dimensional lattice is used for the coding lattice. In the following, we state the coding and shaping lattice properties of our construction, and in the encoding/decoding schemes we detail the construction.

{\bf \noindent Coding lattice: }

The coding lattice $\Lambda_{c,n}$ is an $n$-dimensional lattice with good coding gain, defined by the parity check matrix $\mathbf H$. Let $\mathbf{H}$ satisfy the following conditions. First, $\mathbf{H}$ is lower-triangular. Second, $\mathbf{H}$ is a block matrix, where each block is of size $m \times m$ for $m \ll n$ and $n$ is divisible by $m$, giving $\mathbf{H}$ the following form:

\begin{equation}\label{eqn:generator.structure}
\mathbf {H} = \begin{bmatrix}
      \mathbf H_{11}   & 0 &    0 & 0& 0& 0     \\
      \mathbf H_{21}    & \mathbf H_{22} &    0 & 0& 0& 0   \\
      \mathbf H_{31}    & \mathbf H_{32} &    \mathbf H_{33} & 0& 0& 0 \\
       . & .  &  .   & .& 0 & 0  \\
        .  & . &   .  &.  & . & 0\\
       \mathbf H_{(n/m)1}   & . &  .   & . &  & \mathbf H_{(n/m)(n/m)}
     \end{bmatrix},
\end{equation}
where $\mathbf H_{ij}$ is {a diagonal matrix} for $i=j$ and can have arbitrary structure {for $j < i$}. Let $h_{ij}$ denote the $(i,j)$th element of $\mathbf H$.  {Further, the $r$th block matrix on the diagonal is of the form $h_r \mathbf I_m$, where $\mathbf I_m$ is the $m$-by-$m$ identity matrix.} Let $\bar{\mathbf H}=\mathrm{diag}(\mathbf H)$.

Several lattice families {can be designed to have} this structure, including LDA lattices {and LDLC lattices}. For LDLCs, we can design such a parity check matrix, and because $m$ is small, this constraint has limited impact on the structure of the parity-check matrix for large $n$. For LDLCs, $h_{ii}=1$ and $\mathbf H_{ij}$ has sparse non-zero elements for $i\neq j$.

{\bf \noindent Shaping lattice: }
Let $\Lambda_{s,m}$ be the low-dimensional lattice defined by the generator {matrix} $\Theta \in \mathbb{R}^{m \times m}$, and let $\mathcal V_{s,m}$ be its {fundamental} Voronoi region. Similar to before, $\Theta$ must be lower-triangular and its diagonal entries of $\Theta$ scaled by any diagonal element of $\mathbf H$, defined in \eqref{eqn:generator.structure}, must be an integer, i.e., $h_{jj}\theta_{ii} \in \mathbb Z, \;\; j\in \{0,\ldots n\}$. Finally, {$\Theta$} should satisfy \eqref{eqn:integer.elements}. Let the shaping lattice $\Lambda_{s,n}$ be the $n/m$-fold Cartesian product of $\Lambda_{s,m}$:
\begin{align}\label{eqn:voronoi.product}
\Lambda_{s,n} = \underbrace{\Lambda_{s,m} \times \Lambda_{s,m} \ldots \times \Lambda_{s,m}}_{n/m \; \mathrm{times}}.
\end{align}
Therefore, the Voronoi region of $\Lambda_{s,n}$, denote $\mathcal V_{s,n}$, is the $n/m$-fold Cartesian product of $\mathcal V_{s,m}$.

{\bf \noindent Encoding:} Consider a point-to-point communications channel as described in Sec.~\ref{sec:system} where the source wants to transmit integer information $\mathbf b \in \mathbb Z^n$ to the destination. First, the transmitter divides $\mathbf b$ into $n/m$ blocks, the $r$th block denoted by $\mathbf b^r$, and the $i$th element is selected from following constellation:
\begin{align}
b_{i}^r = \{0,1,\ldots, h_r\theta_{ii}-1\},
\end{align}
where $h_r\theta_{ii}$ is the $i$th diagonal element of the generator matrix $h_r\Theta$, which is related to the scaled shaping lattice $h_r\Lambda_{s,m}$. We define
\begin{align}
\label{eqn:voronoi.mix.info}
\mathbf f^r = \left(\frac{b_{i}^r}{h_r\theta_{11}} \;\; \frac{b_{2}^r}{h_r\theta_{22}} \;\;\ldots \frac{b_{m}^r}{h_r\theta_{mm}}\right)^T,
\end{align}
where $\mathbf f^r \in [0,1)^m$. Then, we map the integer blocks to the fundamental parallelepiped of the scaled shaping lattice $h_r\Lambda_{s,m}$:
\begin{align}
\label{eqn:voronoi.shaped.mix}
\mathbf c^r = h_r\Theta \mathbf f^r,
\end{align}
where $\mathbf c^r \in h_r\mathcal P_m \cap \mathbb Z^m$ by the properties of $\Theta$. Concatenating the result, we obtain an $n$-dimensional integer vector:
\begin{align}
\label{eqn:integer.para}
\mathbf c = [(\mathbf c^1)^T \; \;(\mathbf c^2)^T \ldots (\mathbf c^{n/m})^T ]^T.
\end{align}
Now, similar to before, we create the subtractive dither vector $\mathbf d^r = \Theta \mathbf a^r$ that is uniformly distributed over $\mathcal P_m$. Next, using $n/m$ dither vectors, we form the $n$-dimensional dither vector $\mathbf d = [(\mathbf d^1)^T \; \;(\mathbf d^2)^T \ldots (\mathbf d^{n/m})^T ]^T$.

Then, we subtract the scaled dither vector from $\mathbf c$ to obtain $\mathbf c-\bar{\mathbf H}\mathbf d$. Now, we select an integer vector  $\mathbf k \in \mathbb Z^n$ to satisfy the shaping condition, and we subtract it from $\mathbf c-\bar{\mathbf H}\mathbf d$. The selection of $\mathbf k \in \mathbb Z^n$ is explained later. Next, this vector $\mathbf c-\bar{\mathbf H}\mathbf d-\mathbf k$ is encoded block-wise using the parity check matrix $\mathbf H$. Encoding starts at the first block of $\mathbf c-\bar{\mathbf H}\mathbf d-\mathbf k$ and continues sequentially.  Now, let us consider $r$-th block. Then the $i$th codeword element is:
\begin{align}
\label{eqn:xi}
x'_{i} = \frac{c_{i} -h_{ii}d_i-\sum_{j=1}^{(r-1)m} h_{ij} x_{j}-k_{i}}{h_{ii}},
\end{align}
where $(r-1)m+1\leq i\leq rm$. Note that the summation part in \eqref{eqn:xi} goes only from $1$ to $(r-1)m$ instead of $1$ to $i-1$. This is because the remaining elements from $(r-1)m+1$ to $i-1$ are zero as $\mathbf H_{rr}=h_r\mathbf I_m$.

Next, we form the $m$-dimensional vectors $\mathbf k^r$, $\mathbf x'^r$, $\mathbf c^r$, $\mathbf d^r$, and $\mathbf t^r$ for the $r$th block:
\begin{align}
\mathbf k^r = [k_{(r-1)m+1} \ldots k_{rm}]^T \in \mathbb Z^m,
\end{align}
\begin{align}
\mathbf x'^r = [x'_{(r-1)m+1} \ldots x'_{rm}]^T \in \mathbb R^m,
\end{align}
\begin{align}
\mathbf c^r = [c_{(r-1)m+1} \ldots c_{rm}]^T \in \mathbb Z^m,
\end{align}
\begin{align}
\mathbf d^r = [d_{(r-1)m+1} \ldots d_{rm}]^T \in \mathcal P_m,
\end{align}
and
\begin{align}
\label{eqn:tr}
\mathbf t^r = [\mathbf H_{r1} \mathbf H_{r2} \ldots \mathbf H_{r(r-1)}]\cdot[ (\mathbf x'^1)^T  (\mathbf x'^2)^T \ldots  (\mathbf x'^{r-1})^T]^T.
\end{align}
Based on these definitions, for the $r$th block, we have
\begin{align}
h_r\mathbf x'^r  = \mathbf c^r-h_r\mathbf d^r- \mathbf t^r-\mathbf k^r.
\end{align}
We select $\mathbf k^r$ such that\footnote{Selection of integer $\mathbf k^r$ does not change the $\mathbf t^r$ as $\mathbf H_{rr}$ is a diagonal matrix. For $r=1$, the $\mathbf t_r=\mathbf 0$, hence, $\mathbf k^1$ can be easily found given $\mathbf c^1-h_1\mathbf d^1$. Then, $\mathbf x'^1$ is obtained. Likewise, $\mathbf k^r$ can be sequentially found, before using it to obtain $\mathbf x'^r$.}
\begin{align}
\mathbf k^r &= \mathcal Q_{h_r\Lambda_{s,m}} (\mathbf c^r-h_r\mathbf d_r- \mathbf t^r) \nonumber \\ & = h_r\mathcal Q_{\Lambda_{s,m}} \left(h_r^{-1}(\mathbf c^r-\mathbf d_r- \mathbf t^r)\right).
\end{align}
According to the shaping lattice generator matrix structure, we know $\mathcal Q_{h_r\Lambda_{s,m}}(\cdot) \in \mathbb Z^m$, hence, $\mathbf k^r \in \mathbb Z^m$. Hence,
\begin{align}
\label{eqn:new.rth.block}
\mathbf x'^r &= h_r^{-1}\left[\mathbf c^r-h_r\mathbf d^r- \mathbf t^r-h_r\mathcal Q_{\Lambda_{s,m}} \left(h_r^{-1}(\mathbf c^r-\mathbf d_r- \mathbf t^r)\right)\right]
\nonumber \\ &=\left[h_r^{-1}(\mathbf c^r-h_r\mathbf d^r- \mathbf t^r) -\mathcal Q_{\Lambda_{s,m}} \left(h_r^{-1}(\mathbf c^r-\mathbf d_r- \mathbf t^r)\right)\right]
\nonumber \\ &= [ h_r^{-1}(\mathbf c^r-h_r\mathbf d^r- \mathbf t^r) ] \bmod \Lambda_{s,m}
\nonumber \\ &= [ h_r^{-1}(\mathbf c^r- \mathbf t^r) -\mathbf d^r ] \bmod \Lambda_{s,m}.
\end{align}
The dither $\mathbf d^r$ is uniformly distributed over fundamental parallelepiped $\mathcal P_m$, hence, similar to before, we can show that $[ h_r^{-1}(\mathbf c^r- \mathbf t^r) -\mathbf d^r ] \bmod \Lambda_{s,m}$ is uniformly distributed over the Voronoi region of $\Lambda_{s,m}$, irrespective of the distribution of $\mathbf c^r- \mathbf t^r$. Therefore, $ \mathbf x'^r$ is uniformly distributed over $\mathcal V_{s,m}$, and it has the same average second moment as $\Lambda_{s,m}$ as well as the same NSM.

By concatenating, we find the $n$-dimensional codeword $\mathbf x'$:
\begin{align}
\mathbf x' = [\mathbf x'^1 \; \;\mathbf x'^2 \ldots \mathbf x'^{n/m} ]^T.
\end{align}

As {the} Cartesian product does not change the NSM \cite[Eq. 3.23]{zamir:book14}, we know $\mathbf x'$ has the same NSM as $\Lambda_{s,m}$. Therefore, $\mathbf x'$ has the same shaping gain as $\Lambda_{s,m}$ based on \eqref{eqn:shaping.gain}. The final lattice codeword can be also represented as
\begin{align}
\mathbf x' &= {\mathbf H}^{-1}\left( \mathbf c- \mathbf k - \bar{\mathbf H} \mathbf d\right)
\nonumber \\ & = \mathbf x -{\mathbf H}^{-1}\bar{\mathbf H} \mathbf d,
\end{align}
where $\mathbf x={\mathbf H}^{-1}\left( \mathbf c- \mathbf k\right)$ and $\mathbf k = [(\mathbf k^1)^T \; \;(\mathbf k^2)^T \ldots (\mathbf k^{n/m})^T ]^T \in \mathbb Z^n$.

\remark
The vector $\mathbf t^r$ in \eqref{eqn:tr} {is} related to the high-dimensional coding parity check matrix, and it acts as a self-dither for the quantization step of \eqref{eqn:new.rth.block}. This fact is further investigated in Sec.~\ref{sec:numerical2}.

\remark
The shaping operation, using the $\mathcal M$-algorithm, has complexity $\mathcal O(nd\mathcal M)$, \cite{naftali:ITW09} where $d$ is typically 7 and $\mathcal M$ is the depth of the search ($\mathcal M=151$ was used in the simulations in the following section).   On the other hand, the shaping operation using the $E_8$ shaping lattice, as proposed here, can be accomplished in in about 72 steps \cite[p.~450]{Conway-1999}, so the complexity scales as $72 \frac n m$ (that is $9n$ for the $E_8$ lattice). Both shaping operations are linear in $n$, but for the proposed approach the coefficient on $n$ is lower, and moreover results in better shaping gain.

In the next subsection, we discuss the decoding of our mixed nested lattice code construction.

{\bf \noindent Decoding:} We first show the applicability of this scheme for the AWGN channel. The received signal is given in \eqref{eqn:received}. The first step is to perform lattice decoding using high-dimensional lattice. If a LDLC is used, then the decoder first uses the approximated MAP LDLC decoder in Sec.~\ref{sec:decoding} to obtain the desired integer:
\begin{align}
\mathbf w = {\mathbf H} \mathcal Q_{\Lambda_{c,n}} \left(\frac{\mathbf y+{\mathbf H}^{-1}\bar{\mathbf H} \mathbf d}{r} \right).
\end{align}
Assuming it correctly estimates the integer, $\mathbf w$ is
\begin{align}
\mathbf w = \mathbf c -\mathbf k.
\end{align}
Then we make $n/m$ blocks of them and $r$th block is:
\begin{align}
\mathbf w^r = \mathbf c^r -\mathbf k^r.
\end{align}
Substituting $\mathbf c^r$ in \eqref{eqn:voronoi.shaped.mix}, we have
\begin{align}
\mathbf w^r = h_r\Theta \mathbf f^r -\mathbf k^r,
\end{align}
where $\mathbf k^r \in h_r\Lambda_{s,m}$. Therefore we can represent it as $\mathbf k^r = h_r\Theta \mathbf{\bar k}^r$ where $\mathbf{\bar k}^r \in \mathbb Z^m$. Hence, by substituting these, we obtain
\begin{align}
\mathbf w^r = h_r\Theta \mathbf f^r -h_r\Theta \mathbf{\bar k}^r,
\end{align}
and multiplying by $(h_r\Theta)^{-1}$, we get
\begin{align}
(h_r\Theta)^{-1} \mathbf w^r = \mathbf f^r - \mathbf{\bar k}^r.
\end{align}
Here, $\mathbf f^r$ is the only the fractional part. Hence, $\mathbf f^r$ is
\begin{align}
\mathbf f^r  =  [(h_r\Theta)^{-1}\mathbf w^r] \bmod \mathbb Z^m.
\end{align}
Hence, the desired integer information vector $\mathbf b^r$ can be obtained by
\begin{align}
\mathbf b^r  = h_r\bar \Theta [(h_r\Theta)^{-1}\mathbf w^r] \bmod \mathbb Z^m,
\end{align}
where $\bar \Theta =\mathrm{diag}(\Theta)$. It is possible to use the simple algorithm in Sec.~\ref{sec.voronoi} to perform this modulo operation. As was shown, there always exists an unique solution for $\mathbf b^r$.

\subsection{Algebraic structure for compute-and-forward}
In this subsection, we show that mixed nested lattice codes have the necessary algebraic structure for applications such as compute-and-forward. We prove this by proposing the encoding/decoding steps for the multiple-access compute-and-forward network.

\begin{figure}
\centering
 \includegraphics[width=1\columnwidth]{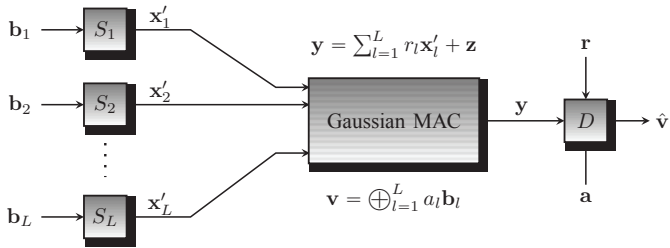}
\caption{System model: Gaussian MAC channel based on compute-and-forward.}
\label{fig:compute.system}
\end{figure}

{\bf \noindent Encoding:} Consider the multiple access channel as shown in Fig.~\ref{fig:compute.system}, which is the basic setting for compute-and-forward.
\label{sec:system.com}
Let $L$ sources use the multiple access channel to simultaneously transmit their signals to a destination. Let $\mathbf b_l \in \mathbb Z^n$ be the integer information, selected from a finite constellation, for the $l$th source. It encodes the integer information to a lattice codeword $\mathbf x'_l\in \Lambda_{c,n}\cap \mathcal V_{s,n}$ using mixed nested lattice encoding as described in Sec.~\ref{sec:new.com}. Each signal $\mathbf x_l$ obeys the same power constraint\footnote{By using different constellation sizes for different sources, it is possible to extend to asymmetric power constraints.}:
\begin{align}
\label{eqn:power.constraint.com}
\frac{1}{n} E\left[ \|\mathbf x'_l\|^2\right] \leq \sigma_{x}^2.
\end{align}
Now all the $L$ sources transmit their power constrained signal $\mathbf x'_l$ via the multiple-access channel as shown in Fig.~\ref{fig:compute.system}. The received signal is
\begin{align}
\mathbf y= \sum_{l=1}^{L} r_l\mathbf x'_l + \mathbf z,
\label{eqn:received.mac}
\end{align}
where $r_l \in \mathbb R$ is the fixed channel coefficient between $l$th user to receiver and $\mathbf z \in \mathbb R^n$ is AWGN noise with per dimension variance $\sigma^2_z$. The receiver is interested in estimating the linear combination
\begin{align}
\mathbf v = \bigoplus_{l=1}^{L} a_{l} \mathbf b_l,
\end{align}
where $a_l$ are integer coefficients and $\bigoplus$ denotes the modulo sum. We represent integer coefficients and channel coefficients in vector form as $\mathbf a^T=[a_1,a_2, \ldots,a_L]$ and $\mathbf r^T=[r_1,r_2, \ldots,r_L]$.

{\bf \noindent Decoding:} First, the destination estimates a linear combination of lattice codewords. For LDLC, this decoder is explained in Appendix \ref{apn:1}. Let us suppose this decoder correctly estimates the linear combination of lattice codewords, given by
\begin{align}
\mathbf u &= \sum_{l=1}^{L} a_{l} \mathbf x_l =\sum_{l=1}^{L} a_{l} \mathbf H^{-1}(\mathbf c_l -\mathbf k_l) = \mathbf H^{-1} \sum_{l=1}^{L} a_{l} (\mathbf c_l -\mathbf k_l).
\end{align}
Multiplying by $\mathbf H$, the destination obtains:
\begin{align}
\mathbf w = \sum_{l=1}^{L} a_{l} (\mathbf c_l -\mathbf k_l).
\end{align}
Then the destination divides $\mathbf w$ into $n/m$ blocks, and the $r$th block is given by
\begin{align}
\mathbf w^r =\sum_{l=1}^{L} a_{l} (\mathbf c_l^r -\mathbf k_l^r).
\end{align}
Substituting for $\mathbf c_l^r$ and $\mathbf k_l^r$ as before, the destination obtains
\begin{align}
\mathbf w^r &=\sum_{l=1}^{L} a_{l} h_r\Theta (\mathbf f_l^r - \mathbf{\bar k}_l^r)  = h_r\Theta \sum_{l=1}^{L} a_{l} (\mathbf f_l^r - \mathbf{\bar k}_l^r).
\end{align}
Now, it multiplies the result by $\Theta^{-1}$ and applies the modulo operation over $\mathbb Z^m$:
\begin{align}
[(h_r\Theta)^{-1} \mathbf w^r] \bmod \mathbb Z^m &= \left[\sum_{l=1}^{L} a_{l} (\mathbf f_l^r  - \mathbf{\bar k}_l^r)\right]\bmod \mathbb Z^m
         \nonumber \\ &= \left[\sum_{l=1}^{L} a_{l} \mathbf f_l^r\right]\bmod \mathbb Z^m.
\end{align}
Finally, it multiplies by $h_r\bar \Theta$:
\begin{align}
\mathbf v^r & = h_r\bar \Theta [(h_r\Theta)^{-1} \mathbf w^r] \bmod \mathbb Z^m \nonumber \\ &= h_r\bar \Theta\left[ \sum_{l=1}^{L} a_{l} \mathbf f_l^r\right]\bmod \mathbb Z^m
        \nonumber \\ & = \left[h_r\bar \Theta \sum_{l=1}^{L} a_{l} \mathbf f_l^r\right]\bmod  h_r\bar \Theta\mathbf 1^m
         \nonumber \\ & = \left[ \sum_{l=1}^{L} a_{l} \mathbf b_l^r\right]\bmod  h_r\bar \Theta\mathbf 1^m.
\end{align}
where $\mathbf 1^m$ is $m$-dimensional all ones vector. By stacking all the $n/m$ blocks, the destination obtains
\begin{align}
\mathbf v &= [\mathbf v^1\;\;\mathbf v^2\ldots \mathbf  v^{n/m}]
   \\ \nonumber   &=\left[\sum_{l=1}^{L} a_{l} \mathbf b_l\right] \bmod \bar {\mathbf H}\bar \Theta^{n}\mathbf 1^n,
\end{align}
where $\bar \Theta^{n} = \mathrm{diag} (\Theta^{n})$ and $\Theta^{n}$ is formed by repeating $ \Theta$ matrix $n/m$ times on the diagonal:
\begin{align}
\Theta^{n} = \begin{bmatrix}
       \Theta   &  &     & & &      \\
          &  \Theta &     & & &    \\
          &  &     & .&  & \\
          &  &     & &  &  \Theta
     \end{bmatrix}.
\end{align}
This proves that we have a mapping $\phi$ that maps $\mathbf b_l \in \mathbb Z^n$, where $i$th element is formed by $b_{li}\in{0,\ldots h_{ii}\theta_{ii}}-1$, to a lattice point in $\Lambda_{c,n} \cap \mathcal V_{s,n}$ such that
\begin{align}
\label{eqn:iso.com}
\phi^{-1}\left(\sum_{l=1}^L a_l \mathbf x_l\right) = \bigoplus_{l=1}^L a_{l}\mathbf b_l,
\end{align}
where $\bigoplus$ is modulo summation over $\bar {\mathbf H}\bar \Theta^{n}\mathbf 1^n$, and
\begin{align}
\label{eqn:phi}
\phi^{-1}\left(\mathbf u\right) = \bar{\mathbf H}\bar\Theta^n \left([(\bar{\mathbf H}\Theta^n)^{-1} \mathbf H \mathbf u]\bmod \mathbb Z^n\right).
\end{align}

\remark
If modulo summation over finite field is required, it is necessary to design generator matrix of the shaping lattice such that $h_{ii}\theta_{ii} = p^{l_i}$ where $p$ is a prime number and $l_i \in \mathbb Z$. For lattices such as scaled $D_m,E_8$ and $BW_{16}$, the related prime number is $p=2$.

\subsection{Numerical evaluation}
\label{sec:numerical2}

\begin{figure}
    \centering
    \includegraphics[width=1\columnwidth]{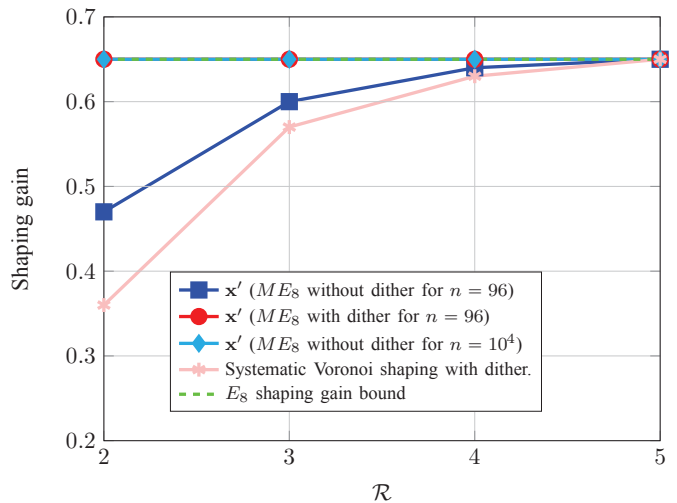}
    \caption{Shaping gain for mixed nested lattice with $E_8$.}
    \label{fig:E8.mixed}
\end{figure}

\begin{figure}
    \centering
    \includegraphics[width=1\columnwidth]{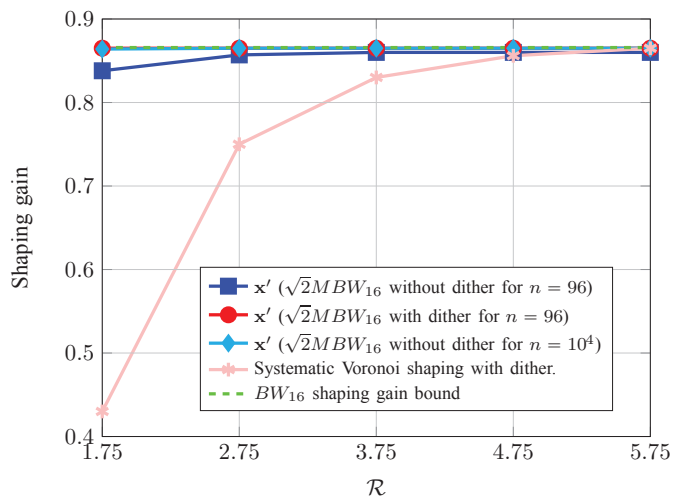}
    \caption{Shaping gain for mixed nested lattice with $BW_{16}$.}
    \label{fig:BW.mixed}
\end{figure}

In this section, we evaluate the performance of the mixed nested lattice code construction. We designed the LDLC parity check matrix based on the properties of \eqref{eqn:generator.structure}. Further, we used the similar degree variations (number of non-zero elements in a row/column) and off diagonal elements of LDLC matrix as in \cite{naftali:ITW09} for our simulations to retain the same coding gains.

Fig.~\ref{fig:E8.mixed} shows the shaping gains of mixed nested lattice construction. First, we have used LDLC with $n=96$ and obtained $0.47, \;0.6,\;0.64$ and $0.65$ dB shaping gains without the dither. However, when we used the random dither (or best fixed dither (\ref{eqn:optimal.dither})), the shaping gain approaches 0.65 dB shaping bound irrespective of the constellation sizes. Then, we have used LDLC with $n=10^4$ and observed shaping gains of $0.65$ dB without using the dither, irrespective of constellation sizes. This is due to the fact that the LDLC codeword component $\mathbf t^r$ in \eqref{eqn:new.rth.block} acts as a self-dither. If we consider LDLC with $n=10^4$, the number of non-zero elements of the majority of the rows/columns is $7$. In this case, each element of $\mathbf t^r$ is a weighted sum of $6$ distinct codeword elements, hence, the number of possible values of $\mathbf t^r$ elements is higher. Therefore, the quantization resolution increases and $\mathbf t^r$ acts as a random dither for larger dimensions. However, for lower-dimensions like $n=96$, the row/column degree is small according to the generator matrix structure \eqref{eqn:generator.structure}, hence, quantization resolution is larger for $\mathbf t^r$ to be a good random dither.

Then, in Fig.~\ref{fig:BW.mixed}, we have simulated the shaping gains of mixed nested lattice using the $BW_{16}$ lattice. We observe similar behavior as with the $E_8$ lattice, and it approaches the $BW_{16}$ shaping bound. Hence, based on Fig.~\ref{fig:BW.mixed} and Fig.~\ref{fig:E8.mixed}, we conclude that mixed nested lattice shaping together with dithering approaches the shaping bound at any constellation size (or any rate). Further, dithering is not necessary for larger constellations or larger LDLC block lengths.

Now, we numerically evaluate a 2-source MAC compute-and-forward network. We have selected the channel vector to be $\mathbf h = [2.1 \;\; 1]^T$. Then the received signal is given by
\begin{align}
\mathbf y = 2.1\mathbf x_1' + \mathbf x_2' + \mathbf z.
\end{align}
The receiver is interested in estimating the linear combination $\mathbf v=a_1 \mathbf x_1 + a_2 \mathbf x_2$. In our simulation, we use the lattice $16\sqrt{2}BW_{16}$ as the shaping lattice to create the mixed nested lattice codebook with LDLC. We have used a coarse constellation in order to protect the symbols that are connected to the last rows of LDLC parity check matrix. This results in slight rate loss ($0.065$ bits/dimension) and the final rates are $\mathcal R_1=\mathcal R_2=4.685$ bits/dim. For the computation rate $\mathcal R_{comp}=4.685$ bits/dim, the respective $\mathrm{SNR}=48.21$ dB, and optimal integer coefficients are $[a_1 \;\; a_2]=[21 \;\; 10]$, which are found using the method proposed in \cite{ordentlich:IT12}.

\begin{figure}
\centering
\includegraphics[width=1\columnwidth]{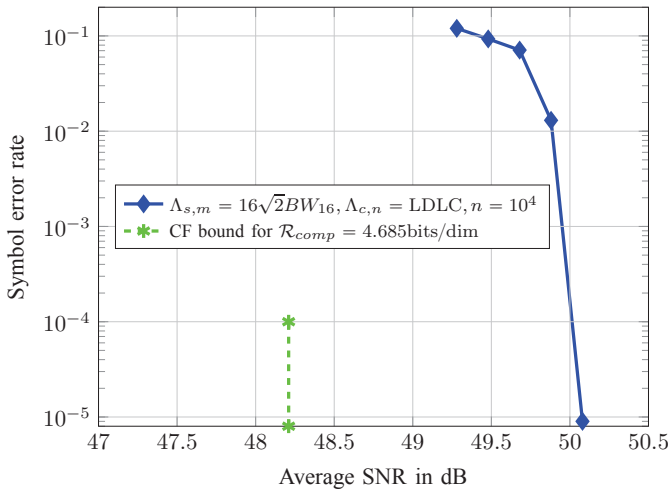}
\caption{Symbol error rate performance of compute-and-forward.}
\label{fig:symbol.com}
\end{figure}

Fig.~\ref{fig:symbol.com} shows the symbol error performance of this setup. It is observed that mixed nested lattice codes are only $1.87$ dB away from the compute-and-forward bound.

\section{Conclusion}

Faced with the need for practical, high-performance generalized shaping methods for lattice codes, we have proposed two new lattice code constructions. By applying low-dimensional, high-gain shaping lattices to short blocks of lattice codewords, we obtain lattice codebooks that have high coding and shaping gains and can be encoded and decoded with low complexity. The first construction can be used for point-to-point AWGN channels to obtain good shaping/coding gains. The second construction {can be used not only for} point-to-point AWGN channels, but also for compute-and-forward scenarios, such as the two-way relay channel, in which a mapping between linear combinations of lattice codes and modulo linear combination of messages is required. We showed our constructions achieve a shaping gain of $0.86$ dB, however, shaping gain of over $1$ dB {should} be easily achieved using the Leech lattice with our methods.

While we have particularized to LDLCs herein, we hasten to point out that the techniques presented in this paper can be applied to any coding lattice with a {check matrix with a } lower-triangular structure, including LDA lattices \cite{dipietro:ISIT13} and one-level LDPC lattices \cite{sadeghi:IWCIT13}. The proposed techniques therefore offer a general step towards the practical realization of the performance advances promised by lattice codes.

\appendix[LDLC decoding for compute and forward.]
\label{apn:1}
The received signal is given in \eqref{eqn:received.mac}. First, the destination adds the dithers. Next, it estimates a linear combination of lattice points:
\begin{align}
\mathbf u = \sum_{l=1}^{L} a_{l} \mathbf x_l,
\end{align}
where $a_{l}$ are an integer coefficients. Simple manipulation shows that $\mathbf u$ is a LDLC lattice point:
\begin{align}
\mathbf u &= \sum_{l=1}^{L} a_{l} \mathbf H^{-1}\mathbf c_l= \mathbf H^{-1} \mathbf c_l',
\end{align}
where $\mathbf c_l'=\sum_{l=1}^{L} a_{l} \mathbf c_l \in \mathbb Z^n$, hence, $\mathbf u$ is a lattice point in the LDLC lattice, \emph{i.e.} $\mathbf u \in \Lambda_{c,n}$ . Further, it is possible to show that $\mathbf u \in \mathcal C'= \sum_{l=1}^{L}a_l \mathcal C$ where $\mathbf x \in \mathcal C$.

Typical compute-and-forward \cite{nazer:IT11} first scales then subtracts the dither to perform lattice decoding. However, here we modify LDLC decoding method to decode to the closest lattice point in the codebook $\mathcal C' = \sum_{l=1}^{L}a_l \mathcal C$. That is equivalent to MAP decoding considering $\mathcal C'$ as the input codebook. The approximated MAP decoder operation is:
\begin{align}
\hat{\mathbf u} &= \arg \max_{\mathbf u}  p(\mathbf u |  \mathbf y )
                \\ \nonumber &=\arg \max_{\mathbf u}  \frac{p(\mathbf y |  \mathbf u )p(\mathbf u ) }{p(\mathbf y)}.
\end{align}
As we want LDLC algorithm to converge to lattice point $\mathbf u$, the input distribution to LDLC algorithm is $p(\mathbf u |  \mathbf y )$. In order to do that we have to find the conditional probability function $p(\mathbf u | \mathbf y)$. First we find the PDF of $p(\mathbf u )$.  We know $\mathbf u$ is a lattice point in LDLC, hence, we use the same trick as in Sec.~\ref{sec:decoding} and make an assumption that the elements of $\mathbf u$ are i.i.d. given $\mathbf u$ is a lattice point to derive the input distribution to LDLC. The distribution is
\begin{align}
p(\mathbf u| \mathbf u \in \Lambda_{c,n}) = \prod_{k=1}^n p(u_k).
\end{align}
Again with the i.i.d assumption, we have
\begin{align}
p(\mathbf y |  \mathbf u ) = \prod_{k=1}^n p(y_k | u_k).
\end{align}
Then we can find
\begin{align}
p(\mathbf u , \mathbf y | \mathbf u \in \Lambda_{c,n} ) &= p(\mathbf y |  \mathbf u )p(\mathbf u| \mathbf u \in \Lambda_{c,n} )
									\\\nonumber & = \prod_{k=1}^n p(y_k | u_k)\prod_{k=0}^n p(u_k)
									\\\nonumber & = \prod_{k=1}^n p(u_k ,  y_k).
\end{align}
Now we need to find the distribution of $p(u_k ,  y_k)$ where $u_k$ and $y_k$ are correlated. We showed that marginal distribution of $\mathbf x_l$ takes approximately a Gaussian distribution, hence, we assume that the marginal distribution of $\mathbf x_l$ given it is a lattice codeword follows is $\mathcal N (0,\sigma_x^2)$. Therefore, with the assumption that $\mathbf x_l$ takes Gaussian distribution, we can prove that $u_k$ has a Gaussian distribution as $u_k$ is sum of independent linear combination of Gaussian random variables, and its distribution is $\mathcal N\left(0,\sum_{l=1}^L a_l^2 \sigma_x^2\right)$. Similarly, based on Gaussian assumptions, $y_k$ has the distribution $\mathcal N \left(0,\sum_{l=1}^L r_l^2 \sigma_x^2+ \sigma_z^2\right)$.\footnote{The assumption of Gaussianity holds for good AWGN coding and shaping lattices as Voronoi region converges to a Gaussian ball. A discussion about this can be found in \cite[Chapter 7]{zamir:book14}. However, for any other lattices they act as approximations and result in shaping and coding losses.}.

With these assumptions, first we find the correlation parameter $\rho_{y_ku_k}$ between $u_k$ and $y_k$:
\begin{align}
\rho_{y_ku_k}&= \frac{\sum_{l=1}^L (r_la_{l})^2 \sigma_x^2 }{\sqrt{\sum_{l=1}^L r_l^2 \sigma_x^2 +\sigma_z^2}\sqrt{\sum_{l=1}^L a_{l}^2x_{l,i}} }
	\\\nonumber &= \frac{ \sigma_x^2(\mathbf r^{T}\mathbf a)^2 }{\sqrt{\sigma_x^2 \|\mathbf r\|^2  +\sigma_z^2}\sqrt{\sigma_x^2 \| \mathbf a\|^2} }.
\end{align}
Then, we can rewrite the variances of $y_k$ and $u_k$ in vector form as $\sigma_y^2=\sigma_x^2 \|\mathbf r\|^2  +\sigma_z^2$ and  $\sigma_u^2=\sigma_x^2 \| \mathbf a\|^2$.
Then we can find
\begin{align}
\label{eqn.compute.distribution}
p(u_k |  y_k) = &\frac{1}{\sqrt{2\pi \sigma_x^2\left(\sigma_z^2||\mathbf a||^2-\frac{\sigma_x^2(\mathbf r^T\mathbf a)^2}{\sigma_z^2+\sigma_x^2||\mathbf r||^2}\right)}} \nonumber \\ & \exp\left({\frac{-\left (u_k-\frac{(\mathbf r^T\mathbf a) y_k}{\sigma_z^2+\sigma_x^2||\mathbf r||^2}\right)^2}{2\sigma_x^2\left(\sigma_z^2||\mathbf a||^2-\frac{\sigma_x^2(\mathbf r^T\mathbf a)^2}{\sigma_z^2+\sigma_x^2||\mathbf r||^2}\right)}}\right).
\end{align}

Hence, we use \eqref{eqn.compute.distribution} as the input distribution to the $k^{th}$ variable node of LDLC algorithm. Using this decoder, the destination obtains an estimate for $\mathbf u$.
\remark
Based on the input distribution to the decoder, we can show that this LDLC decoder converges to the compute-and-forward results \cite{nazer:IT11} and it gets to the same computation rates apart from the rate losses of practical codes.
\bibliographystyle{IEEEtran}
\bibliography{arxiv_upload}

\end{document}